\documentclass[epj]{svjour}
\usepackage{graphics}
\begin{document}
\title{Effect of screening of the electron-phonon interaction on mass renormalization and optical conductivity of the Extended Holstein model polarons}
\author{B. Ya. Yavidov
}                     
%
%
\institute{Institute of Nuclear Physics, 100214 Ulughbek, Tashkent, Uzbekistan}
%
\date{Received: date / Revised version: date}
%
\abstract{
An interacting electron-phonon system is considered within the Extended Holstein model at strong coupling regime and nonadiabatic approximation.
It is assumed that screening of an electron-phonon interaction is due to the excess electrons in a lattice. An influence of the screening on the mass and optical conductivity of a lattice polarons is studied.  A more general form Yukawa-type electron-phonon interaction potential potential is
accepted and corresponding forces are derived in a lattice. It is emphasized that the screening effect is more pronounced at the values of screening
radius comparable with a lattice constant. It is shown that the mass of a lattice polaron obtained using Yukawa-type electron-phonon interaction
potential is less renormalized than those of the early studied works at the same screening regime. Optical conductivity of lattice polarons
 is calculated at different screening regimes. The screening lowers the value of energy that corresponds to the peak of the optical conductivity curve. The shift (lowering) is more pronounced at small values of screening radius too. The factors that give rise to this shift is briefly discussed.
\PACS{
      {71.38.-k} {Polarons and electron-phonon interactions},
      {71.38.Ht} {Self-trapped or small polarons,}
      {78.67.-n} {Optical properties of low-dimensional, mesoscopic, and nanoscale materials and structures}
     } 
} 
\authorrunning{B. Ya. Yavidov}
\titlerunning{Effect of screening of the electron-phonon interaction\ldots}
\maketitle
\section{Introduction}
A sufficiently strong interaction of a charge carrier with vibration of a lattice (phonon field) could give rise to a self-trapping phenomenon or {\it polaron} formation in solids. The possibility of such a phenomenon in alkali-halide solids was first predicted by Landau \cite{land}. Since then studies on polaron physics have been made by a number of scientists both theoretically and experimentally. The theoretical research up to date continues within the major frameworks: (i) Fr\"{o}hlich model \cite{fro2}, (ii) molecular-crystal Holstein model (HM) \cite{hol} and etc. In the first model polaron forms due to an interaction of an electron with the longitudinal optical vibrations of a polar ionic crystalline. The crystal assumes as a continuum. This means one neglects detail structure of a lattice. In the second case polaron formation is due to coupling of a charge carrier to a intramolecular vibrations of a lattice. Holstein model is commonly studied in a discrete lattice. In the last decades both models have been extensively studied (see for example reviewer papers \cite{shl-ston,amr,asa-devr} and books \cite{firsov,bry}).

A model of a polaron with a long-range "density -displacement" type interaction was introduced  in Ref. \cite{alekor} by Alexandrov and Kornilovitch.
The model by itself represents an extension of the Fr\"{o}hlich polaron model \cite{fro2} to a discrete ionic crystal lattice or extension of the Holstein polaron model \cite{hol} to the case when an electron interacts with many ions of a lattice with longer ranged electron-phonon interaction force. Subsequently, the model was named as the extended Holstein model (EHM) \cite{flw}. The model \cite{alekor} was introduced in order to mimic  $high-T_{c}$ cuprates, where the in-plane (CuO$_2$) carriers are strongly coupled to the $c$-axis polarized vibrations of the $apical$ oxygen ions \cite{timusk}.
Results for the mass of EHM polaron with the only $z$- polarized vibrations of ions were obtained (i) analytically in strong coupling regime and nonadiabatic limit and (ii) numerically in intermediate coupling regime and near-nonadiabatic limit with the help of Quantum Monte-Carlo method. It was established that at strong coupling regime $\lambda\gg 1$ ($\lambda=E_p/zt$, where $E_p$ is polaron shift, $z$ is lattice coordination number and $t$ is the nearest neighbor hopping integral) mass of EHM polaron is less renormalized than mass of ordinary Holstein model polaron. In the opposite regime mass of EHM polaron is more renormalized than mass of ordinary Holstein model polaron.
Conclusions of Ref.\cite{alekor} concerning mass renormalization were confirmed later by the other authors \cite{flw,bt,trg}. Fehske, Loos and Wellein \cite{flw} investigated the electron-lattice correlations, single-particle spectral function and optical
conductivity of the EHM in the strong and weak coupling regimes by means of an  exact Lancroz diagonalization method. Bon\v{c}a and
Trugman in Ref. \cite{bt} and Trugman, Bon\v{c}a and Li-Chung Ku in Reg. \cite{trg} studied the EHM on different lattices and showed that
ions' arrangement has significant influence on mass renormalization. Moreover Trugman, Bon\v{c}a and Li-Chung Ku discussed the
influence of $y$- polarized vibrations of ions to the mass renormalization and found that this type vibrations give rise to more renormalization of polaron mass than $z$- polarized vibrations of ions.  Two-site system of the model in strong coupling regime and extreme adiabatic limit was studied in \cite{asaya,yavjetp,yavphysb}. Other properties of EHM such as the ground state dispersion, the density of states, the ground state spectral weight, the average kinetic energy and the mean number of phonons by means of the variational and Quantum Monte Carlo simulation approaches have been studied in Refs.\cite{kor-prb60,pcf,hel}.
A more realistic case, when the {\it apical} ions are three-dimensional anisotropic oscillators, is considered in
\cite{asakor} within the framework of   Fr\"{o}hlich-Coulomb model in the strong coupling and nonadiabatic limit. The EHM with a screened electron-phonon interaction was discussed in Refs.\cite{kor-giant,spencer,hague-etal,hague-kor}.

Several modified versions of Holstein model were applied to study semiconducting polymers \cite{stojan-bobbert,meisel-bobbert}, and charge transport in disordered systems \cite{tri-fri,tri} and DNA molecules \cite{tri-sim-kar}. The works \cite{stojan-bobbert,meisel-bobbert} treat an electron-phonon interaction as nonlocal and based on a variational approach. Small polaron transport through DNA molecules was investigated within the framework of generalized molecular crystal model (GMCM) in Ref. \cite{tri-sim-kar} in which electronic energy and electron-phonon coupling constant are different for different sites.

At the same time polarons were experimentally recognized as quasiparticles in the novel materials, in particular, in the superconducting
cuprates and colossal magnetoresistance manganites \cite{bar-bishop,asa-advanced-mat}.

In this paper a particular question of coupled electron-phonon system will be considered within the EHM.
Namely, it is an influence of a screening of an electron-lattice interaction on the mass and the optical conductivity of lattice polarons. In contrast to Refs.\cite{kor-giant,spencer,hague-etal,hague-kor} an explicit form of electron-lattice interaction forces will be derived associated with the different type of the polarized vibrations and the optical conductivity of the EHM polarons at different values of screening
radius will be presented. We will see that an effect of the screening is more pronounced at small values of the screening radius and briefly discuss
applicability of the early applied screened electron-phonon forces at short distances.
\begin{figure}
\resizebox{0.5\textwidth}{!}{%
  \includegraphics{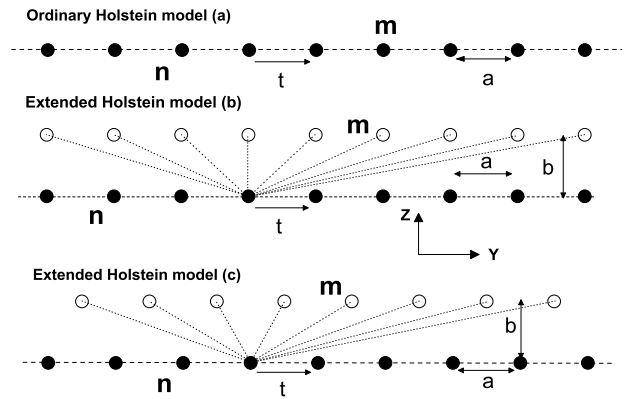}
}
\caption{(a) In ordinary Holstein model an electron moves in a one-dimensional chain of the molecules and interacts with the single site intramolecular vibrations. (b and c) In the extended Holstein model an electron hops on a lower chain and
interacts with the ions vibrations of an upper infinite chain via a density-displacement type force $f_{{\bf m},\alpha}(\bf n)$.
Two lattices (b) and (c) are differ from each another by the shift of upper (or lower) chain to a distance $a/2$ along $y-$ direction.
The distances between the chains ($b$) and between the ions of the same chain ($a$) are assumed equal to 1.
Dotted lines represents an interaction of an electron on site ${\bf n}$ with the ions of the upper chain.}
\label{fig:1}       
\end{figure}
\section{Hamiltonian and screened force}
We consider an electron performing hopping motion on a lower chain consisting of the static sites, but interacting with all ions of an upper chain via a long-range density-displacement type force, as shown in Fig.1(b) and Fig.1(c). So, the motion of an electron is always one-dimensional, but a vibration of the upper chain's ions is isotropic and two-dimensional one. Authors of Ref. \cite{alekor} studied a polaron formation and its mass renormalization within the framework of a quite general electron-phonon model:
\begin{equation}\label{1}
H=H_{e}+H_{ph}+H_{e-ph},
\end{equation}
where
\begin{equation}\label{2}
H_{e}=-t \sum_{\bf n}(c^{\dagger}_{\bf n}c_{\bf n+a}
+H.c.)
\end{equation}
is the electron hopping energy,
\begin{equation}\label{3}
H_{ph}=\sum_{{\bf m},\alpha}\left(-\frac{\hbar^2\partial^2}{2M\partial u^{2}_{{\bf m},\alpha}}+\frac{M\omega^2u^{2}_{{\bf m},\alpha}}{2}\right)
\end{equation}
is the Hamiltonian of the vibrating ions,
\begin{equation}\label{4}
H_{e-ph}=\sum_{{\bf n,m},\alpha}f_{{\bf m},\alpha}({\bf n})\cdot u_{{\bf m},\alpha}c^{\dagger}_{\bf n}c_{\bf n}
\end{equation}
describes interaction between the electron which belongs to the lower chain and the ions of the upper chain. Here $c^{\dagger}_{\bf n}$($c_{\bf n}$) is the creation (destruction) operator of an electron on the site $\bf n$, $u_{{\bf m},\alpha}$ is the $\alpha=y,z$- polarized displacement of the {\bf m}-th ion and $f_{{\bf m},\alpha}({\bf n})$ is an interacting density-displacement type
force between an electron on the site {\bf n} and the $\alpha$ polarized vibration of the {\bf m}-th ion. $M$ is the mass of the vibrating ions
and $\omega$ is their frequency. A case of an electron coupled to single site intramolecular vibrations represents the canonical Holstein model (Fig.1(a)) in which electron-phonon interacting force is defined as $f_{{\bf m},\alpha}(\bf n)=\kappa_\alpha\delta_{{\bf m},{\bf n}}$. The early studies of EHM \cite{alekor,flw,bt,trg,asaya,yavjetp,yavphysb,kor-prb60,pcf} were performed with the unscreened electron-phonon interaction forces (see below Eq.(6) and Eq.(7) at $R=\infty$). The force was deduced from pure Coulomb potential $\sim const/r$ which for our discrete lattice can be written as $\sim const/\sqrt{|{\bf n}-{\bf m}|^2+b^2}$. The distance along the chain $|{\bf n}-{\bf m}|$ is measured in units of lattice constant $|{\bf a}|=1$. The distance between the chains is $|{\bf b}|=1$ too. Detailed derivation of the unscreened force can be found in Ref.\cite{spencer}. EHM with screened electron-lattice interaction force was studied in Refs.\cite{kor-giant,spencer,hague-etal,hague-kor}. Is was assumed that screening of the electron-lattice interaction is due to the presence of other electrons in the lattice. However, an explicit derivation of the screened force was not presented.

In general, at present there is no exact analytical expression for a screened electron-ion force. Commonly used formulas for screened forces are obtained under some approximations. Here we consider a more general form of electron-lattice interaction force due to the direct Coulomb forces of an electron in the lower chain with ions of the upper chain. Namely, we approximate an electron-ion interaction potential as the Yukawa potential:\\ $\sim const\exp[-r/R]/r$, where $R$ is the screening radius and $r$ is the position radius. Such type of approximation is more suitable to the real systems, in particular, to cuprates. Indeed, cuprates change their properties upon doping from insulating state to metallic one. In such circumstances, the choice of Yukawa potential seems to be appropriate since one has to consider different doping regimes. In optimally and overdoped regimes, cuprates are believed to be in a metallic state and one expects the form of electron-ion potential would be $(Ze^2/r)\exp{[-r/R_{TF}]}$, where $R_{TF}=(E_F/2\pi e^2n_0)^{1/2}$ is the Thomas-Fermi screening radius, $E_F$ is the Fermi energy, $n_0$ is an equilibrium charge density. In the opposite underdoped regime cuprates are likely in a semiconducting state and thus one can use Debye approximation for the screened electron-ion potential\\ $(Ze^2/r)\exp{[-r/R_D]}$, where $R_D=(\varepsilon_0 k_BT/4\pi e^2n_0)^{1/2}$ is the Debye screening radius, $T$ is an absolute temperature, $\varepsilon_0$ is static dielectric constant of cuprates and $k_B$ is the Boltzmanm constant. Since in the both regimes electron-ion potentials have exponential term our choice seems to be more realistic. Then discrete form of the electron-ion potential is written as:
\begin{eqnarray}
  U_{{\bf m}}({\bf n}) &=& \frac{\kappa}{(|{\bf n}-{\bf m}|^2+b^2)^{1/2}}\times \\
\nonumber   &\times& \exp\left[-\frac{\sqrt{|{\bf n}-{\bf m}|^2+b^2}}{R}\right],
\end{eqnarray}
where $\kappa$ is some coefficient and $R$ is measured in units of $|\bf a|$. From the potential Eq.(5) one obtains an analytical expressions for the $z$- and $y$- type components of the screened electron-lattice forces:
\begin{eqnarray}
\nonumber  f_{{\bf m},y}({\bf n}) &=& \frac{\kappa |{\bf n}-{\bf m}|}{(|{\bf n}-{\bf m}|^2+b^2)^{3/2}} \left(1+\frac{\sqrt{|{\bf n}-{\bf m}|^2+b^2}}{R}\right) \\
   &\times& \exp\left[-\frac{\sqrt{|{\bf n}-{\bf m}|^2+b^2}}{R}\right]
\end{eqnarray}
and
\begin{eqnarray}
\nonumber  f_{{\bf m},z}({\bf n}) &=& \frac{\kappa b}{(|{\bf n}-{\bf m}|^2+b^2)^{3/2}} \left(1+\frac{\sqrt{|{\bf n}-{\bf m}|^2+b^2}}{R}\right) \\
   &\times& \exp\left[-\frac{\sqrt{|{\bf n}-{\bf m}|^2+b^2}}{R}\right].
\end{eqnarray}
\begin{figure}
\resizebox{0.5\textwidth}{!}{%
  \includegraphics{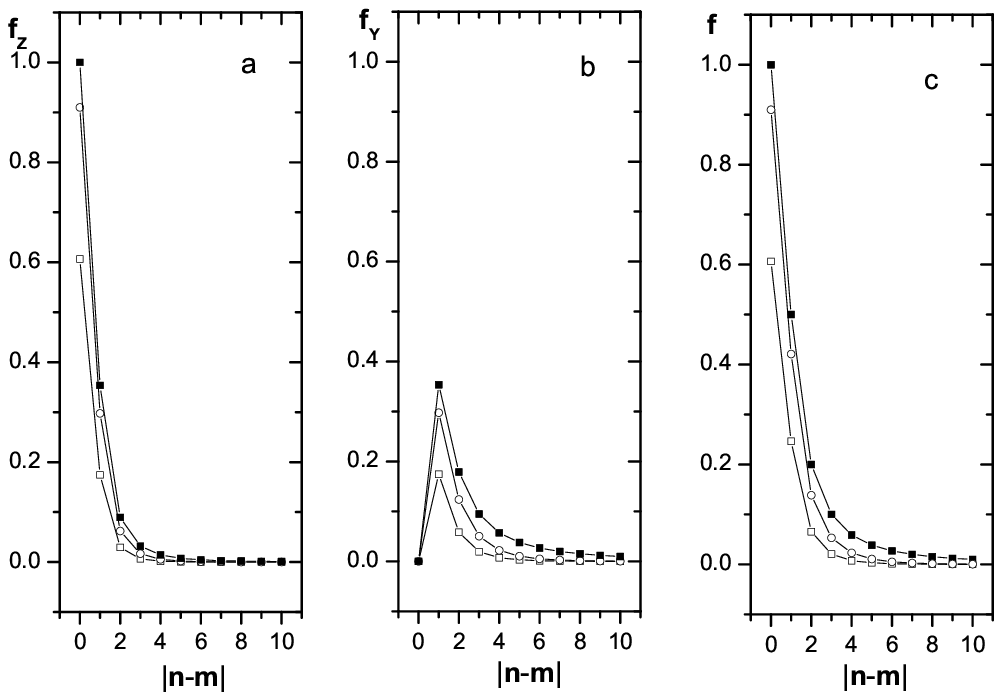}
}
\caption{The values of electron-phonon interaction forces as a function of $|{\bf n}-{\bf m}|$. Filled squares, open squares and open circles correspond to unscreened, screened according to Refs.\cite{kor-giant,spencer,hague-etal,hague-kor} and to our
case Eq.(6) and Eq.(7), respectively. Forces are in units of $\kappa $ and calculated for the lattice in Fig.1(b) at $R=2$.}
\label{fig:2}       
\end{figure}
\begin{figure}
\resizebox{0.5\textwidth}{!}{%
  \includegraphics{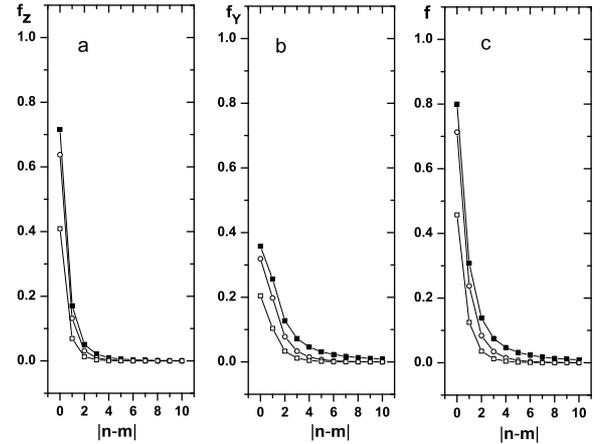}
}
\caption{The values of electron-phonon interaction forces as a function of $|{\bf n}-{\bf m}|$. Filled squares, open squares and open circles correspond to unscreened, screened according to Refs.\cite{kor-giant,spencer,hague-etal,hague-kor} and to our
case Eq.(6) and Eq.(7), respectively. Forces are in units of $\kappa $ and calculated for the lattice in Fig.1(c) at $R=2$.}
\label{fig:3}       
\end{figure}
The dependencies of the electron-lattice interaction forces $f_{{\bf m},z}({\bf n})$, $f_{{\bf m},y}({\bf n})$ and the full force $f_{{\bf m}}({\bf n})=\\\sqrt{f^2_{{\bf m},z}({\bf n})+f^2_{{\bf m},y}({\bf n})}$ on distance $|{\bf n}-{\bf m}|$ for the lattices of Fig.1(b) and Fig.1(c) are presented in Fig.2 and Fig.3, respectively. In the same figures unscreened force of Ref.\cite{alekor} and screened force of Refs.\cite{kor-giant,spencer,hague-etal,hague-kor} are also plotted. Comparison of the forces show that at large distances $|{\bf n}-{\bf m}|$ the differences between the forces are small.
However, at small $|{\bf n}-{\bf m}|$ the forces are strongly deviate from each other. Here the short distances are of considerable interest since:
(i) doping of cuprates reduces screening radius $R$ and, (ii) therefore, the polaronic effects come only from neighboring ions. The importance of
nearest neighbor ions  and their arrangement in determining polaron parameters was already recognized \cite{bt,trg}. Interesting point is that $y$- component of the electron-lattice interaction force $f_{{\bf m},y}({\bf n})$ has no effect to an electron at ${\bf m}={\bf n}$ when considering lattice in Fig.1(b). This is not the case in lattice Fig.1(c) where ions of the upper chain are shifted along $y$- direction to distance $a/2$ \cite{bt,trg} (see Fig.3b). As one can see from Eq.(7) the screened electron-phonon interaction force used in Refs.\cite{kor-giant,spencer,hague-etal,hague-kor} (see for example Eq.(16) of Ref.\cite{spencer}) is a particular case of the more general type of force induced by the Yukawa type potential Eq.(5). Indeed, if one assumes validity of the conditions $|{\bf b}|\ll |{\bf n}-{\bf m}|\ll R$, our Eq.(7) reduces to Eq.(16) of Ref.\cite{spencer}.
Thus the early studied screened electron-phonon interaction force represents a particular case of Eq.(7) at $R\gg |{\bf n}-{\bf m}|$. Unscreened force of Ref.\cite{alekor} ($z$- component)and Ref.\cite{bt} may be considered as the particular cases of our force at $R=\infty$ and $R=1$, respectively.

\section{Mass renormalization}
Let's discuss an effect of screening radius on the mass of EHM polaron at strong coupling regime $\lambda\gg 1$ and nonadiabatic limit $t/\hbar\omega<1$. One can see the main features of the EHM are: (i) the electronic (2), the phonon (3) and the electron-phonon (4) Hamiltonians all have the same periodicity $a$, (ii) the electron-phonon interaction density-displacement type forces following Eq.(5) and Eq.(6) are translational invariant , i.e. $f_{{\bf m},\alpha}({\bf n})=f_{{\bf m}-{\bf n}',\alpha}({\bf n}-{\bf n}')$. In Ref.\cite{asaya} a simple two-site model of a small polaron within the framework of EHM was studied in the strong coupling limit. The features (i) and (ii) of the model enables us to apply Bloch theorem  and extend the results of the two-site model \cite{asaya} to our lattice (Fig.1(b) and Fig.1(c)) and obtain polaronic band as
\begin{equation}\label{8}
E(k)=N\hbar\omega-E_{p}-2\tilde{t}\cos(ka),
\end{equation}
where $N$ is a number of ions in the upper chain,
\begin{equation}\label{9}
    E_{p}=E_{p}({\bf n})=\sum_{{\bf m},\alpha}\frac{f^{2}_{{\bf m},\alpha}({\bf n})}{2M\omega^2}
\end{equation}
is the polaronic shift which is independent of ${\bf n}$,
\begin{equation}\label{10}
\widetilde{t}=te^{-g^{2}}
\end{equation}
is a renormalized hopping integral and
\begin{equation}\label{11}
g^2=\frac{1}{2M\hbar\omega^3}\sum_{{\bf m},\alpha}[f^2_{{\bf m},\alpha}({\bf 0})-f_{{\bf m},\alpha}({\bf 0})f_{{\bf m},\alpha}({\bf 1})].
\end{equation}
The same result may be obtained using an analytical method based on the extended (or nonlocal) Lang-Firsov transformation and subsequently perturbation theory with respect to the parameter $1/\lambda$ \cite{alekor}. The EHM polaron mass can be expressed in terms of electron-phonon coupling constant as (in units of the bare band mass)
\begin{equation}\label{12}
    m_p=e^{2\lambda\gamma t/\hbar\omega},
\end{equation}
where
\begin{equation}\label{13}
    \gamma=1-\frac{\sum_{{\bf m},\alpha}f_{{\bf m},\alpha}({\bf 0})\cdot f_{{\bf m},\alpha}({\bf 1})}{\sum_{{\bf m},\alpha} f_{{\bf m},\alpha}^2({\bf 0})}.
\end{equation}
As one can see from Eq.(13) $\gamma$ it is independent of $\kappa$, but depends on the geometry of the lattice and on the range of force.  For the ordinary Holstein model with local interaction it is always equal to 1. However for EHM with the long-range interaction forces Eq.(6) and Eq.(7) $\gamma$ is smaller than one.
\begin{figure}
\resizebox{0.5\textwidth}{!}{%
  \includegraphics{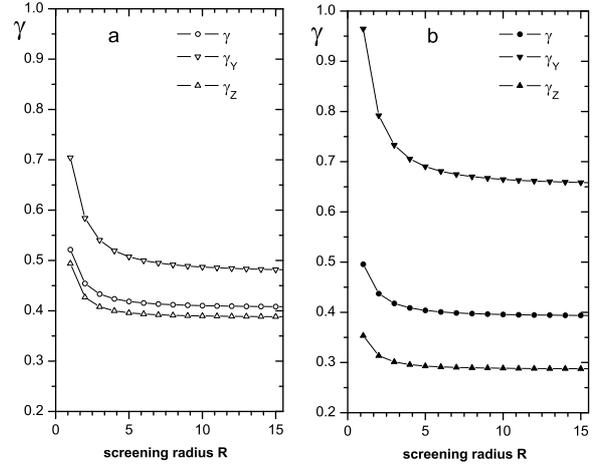}
}
\caption{Plot of $\gamma$ as a function of screening radius $R$ for the lattices: a - Fig1.(b) and b - Fig.1(c).}
\label{fig:4}       
\end{figure}
\begin{table}
\caption{\label{tab:table1} The calculated values of $\gamma$ for the lattices Fig.1(b) and Fig.1(c) at different $b$.}
\begin{center}
\begin{tabular}{|c|ccc|}
\hline
$R$&$b=a/2$&$b=a$&$b=2a$\\
\hline
&&for Fig.1(b)&\\
\hline
       1&0.8626&0.5213&0.2217\\
       2&0.8310&0.4543&0.1704\\
       3&0.8214&0.4332&0.1532\\
       4&0.8173&0.4237&0.1449\\
       5&0.8151&0.4185&0.1403\\
$\infty$&0.8104&0.4064&0.1280\\
\hline
&&for Fig.1(c)&\\
\hline
       1&0.8856&0.4957&0.2210\\
       2&0.8422&0.4372&0.1701\\
       3&0.8278&0.4176&0.1530\\
       4&0.8213&0.4086&0.1448\\
       5&0.8179&0.4036&0.1402\\
$\infty$&0.8101&0.3920&0.1279\\
\hline
\end{tabular}
\end{center}
\end{table}

\begin{figure}
\resizebox{0.5\textwidth}{!}{%
  \includegraphics{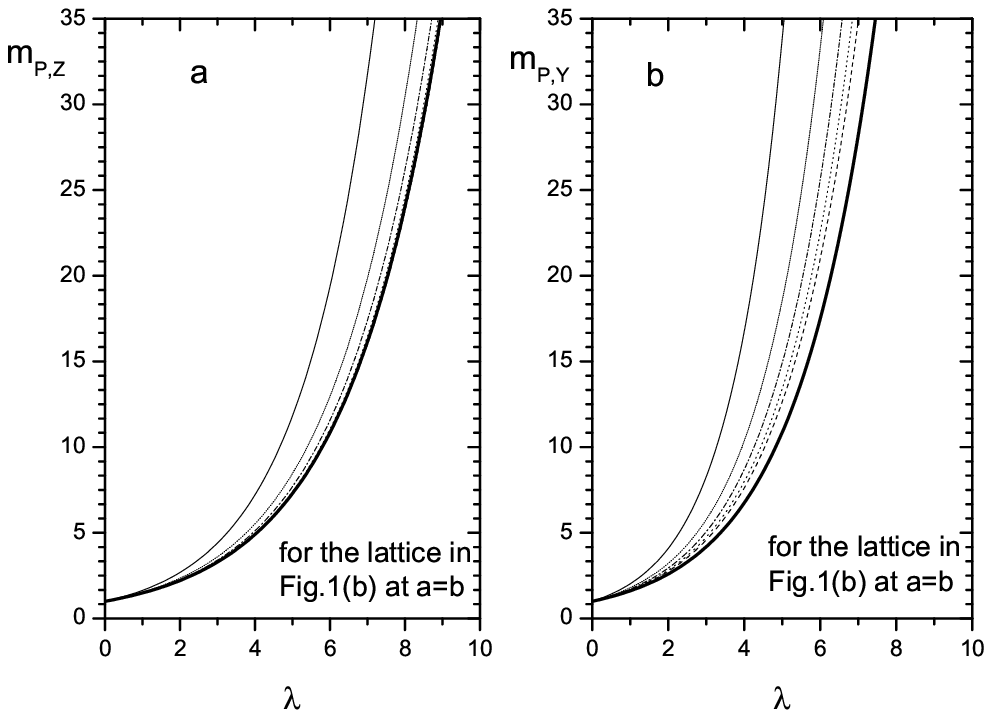}
}
\caption{Polaron masses $m_{p,z}$ (a) and $m_{p,y}$ (b) as a function of electron-phonon coupling constant $\lambda$  at different values of screening radius $R$: $R=1$- thin line, $R=2$- short-dotted line, $R=3$- dash-dotted line, $R=4$- dotted line,
 $R=5$- dashed line and $R=\infty$- solid thick line. $t/\hbar\omega=0.5$.}
\label{fig:5}       
\end{figure}
\begin{figure}
\resizebox{0.5\textwidth}{!}{%
  \includegraphics{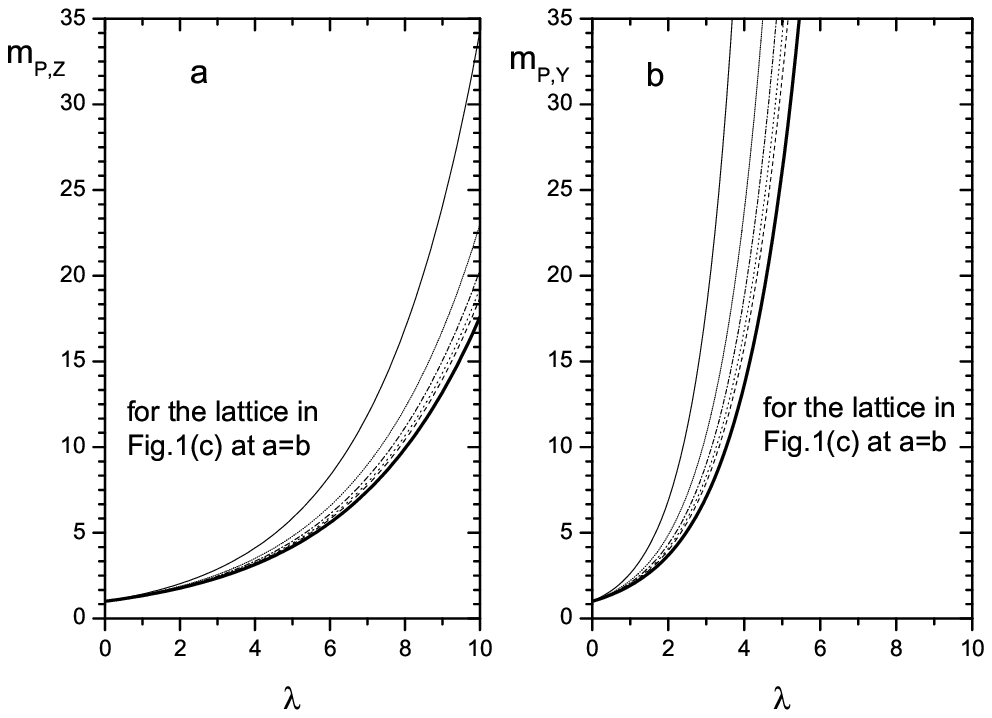}
}
\caption{Polaron masses $m_{p,z}$ (a) and $m_{p,y}$ (b) as a function of electron-phonon coupling constant $\lambda$  at different values of screening radius $R$: $R=1$- thin line, $R=2$- short-dotted line, $R=3$- dash-dotted line, $R=4$- dotted line,
 $R=5$- dashed line and $R=\infty$- solid thick line. $t/\hbar\omega=0.5$.}
\label{fig:6}       
\end{figure}
\begin{figure}
\resizebox{0.5\textwidth}{!}{%
  \includegraphics{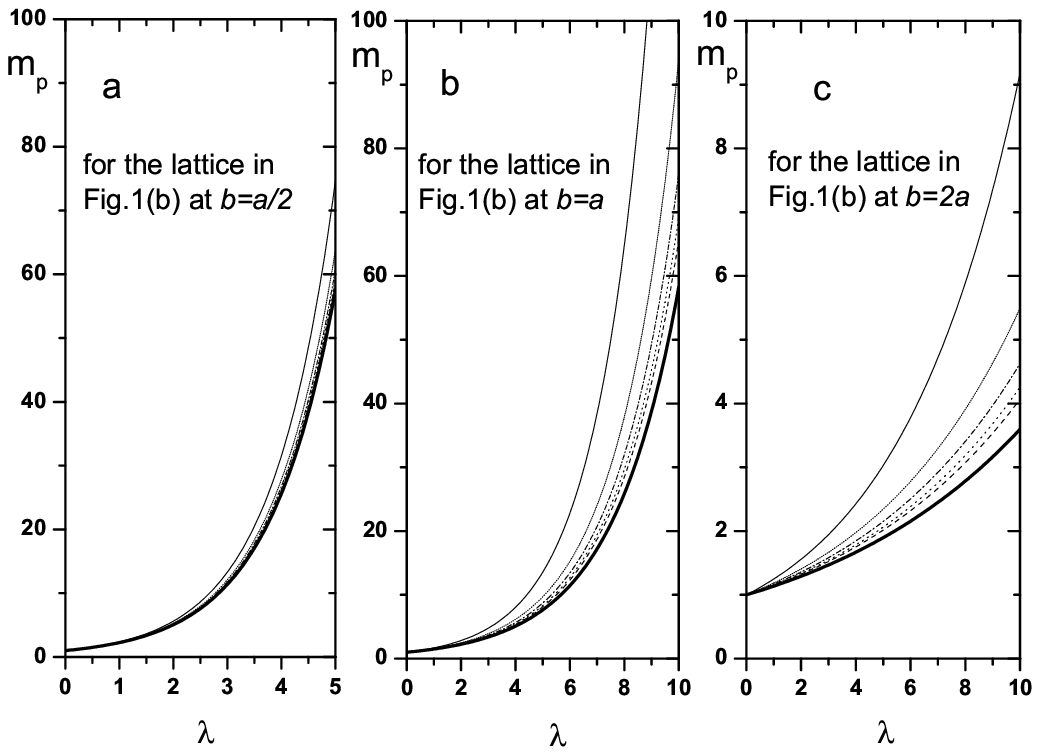}
}
\caption{The full polaron mass $m_p$ as a function of electron-phonon coupling constant $\lambda$  at different values of screening radius $R$ ($R=1$- thin line, $R=2$- short-dotted line, $R=3$- dash-dotted line, $R=4$- dotted line,
 $R=5$- dashed line and $R=\infty$- solid thick line) and ratio $a/b$. $t/\hbar\omega=0.5$.}
\label{fig:7}       
\end{figure}
\begin{figure}
\resizebox{0.5\textwidth}{!}{%
  \includegraphics{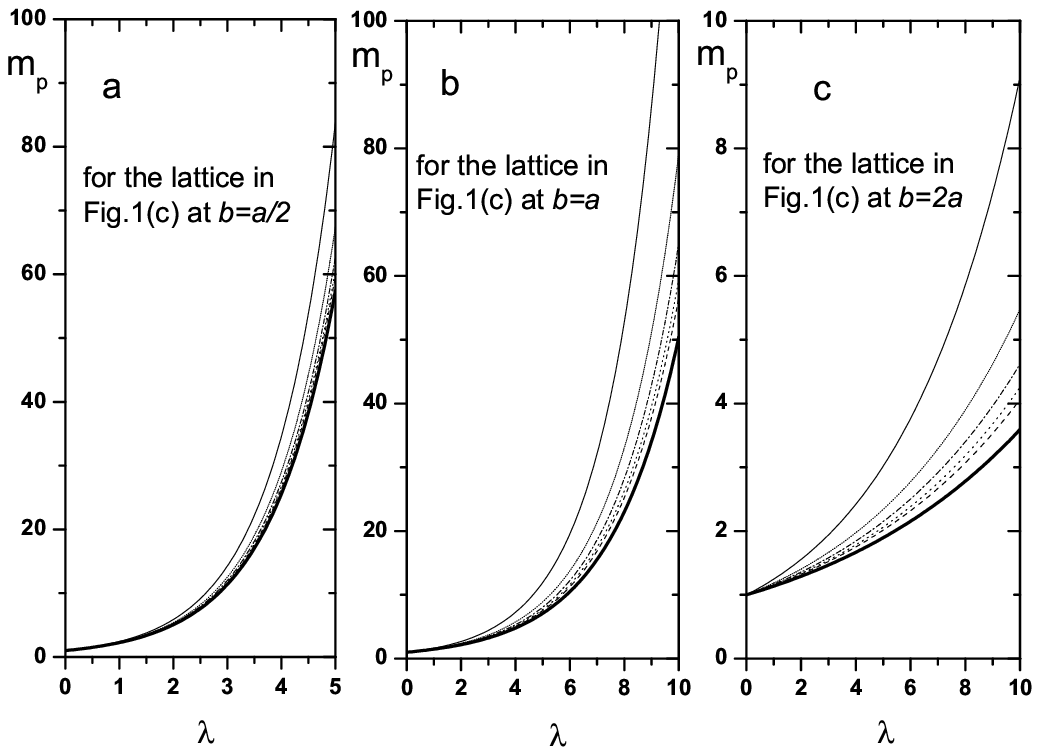}
}
\caption{The full polaron mass $m_p$ as a function of electron-phonon coupling constant $\lambda$  at different values of screening radius $R$ ($R=1$- thin line, $R=2$- short-dotted line, $R=3$- dash-dotted line, $R=4$- dotted line,
 $R=5$- dashed line and $R=\infty$- solid thick line) and ratio $a/b$. $t/\hbar\omega=0.5$.}
\label{fig:8}       
\end{figure}
As it is seen from Eqs.(8)-(13), all polaron parameters are affected by $R$ as it enters to all expressions. We have considered a situation when a polaron is formed by (i) the only $z$- ($y$-) polarized vibrations of the upper chain and (ii) the both $z$- and $y$- polarized vibrations of the upper chain. For each situation we have calculated $\gamma_z$ ($\gamma_y$) and $\gamma$ for the lattices under consideration at different $R$. The results are presented in Fig.4. When the screening radius $R$ is large enough compared to the lattice constant $|{\bf a}|$, the effect of screening on $\gamma$ is small. However, as the screening radius $R$ is decreased, the effect becomes more sensitive. The general tendency is that all of them decreases with the screening radius $R$. Plot of polaron masses $m_{p,z}$ and $m_{p,y}$ corresponding to $\gamma_z$ and $\gamma_y$ as a function of $\lambda$ are given in Fig. 5 and Fig.6, respectively. A value of $\gamma$ which determines mass of EHM polaron is sensitive to the ratio $b/a$. The calculated values of $\gamma$ (when both $z$- and $y$- polarized vibrations are taken into account) at different ratio $b/a$ are presented in Table 1. We have also calculated the full mass of EHM polaron at different $b/a$. The results are presented in Fig.7 and Fig.8. As one can see shift of the whole upper chain in $z-$ direction to some distance has crucial impact to mass renormalization. Meanwhile, shift of the upper chain to distance $a/2$ along $y-$ direction don't give rise the strong change of the full polaron mass. We confirm early findings which indicate that the unscreened electron-phonon interaction provides a more mobile polaron and the polaron with the screened electron-phonon interction has a more renormalized mass \cite{spencer}. At the same time our study differs from the early studied works by the form of electron-phonon force. In our study the screened electron-lattice forces are derived with the help of more general form of electron-ion interaction potential, which is Yukawa potential, and can be microscopically derived (see for example \cite{anselm,mahan}). When screening radius is comparable to the lattice constant our force strongly deviates from the early studied force. It lies somewhere in the middle of the totally unscreened force of Ref.\cite{alekor} and the screened force of Ref.\cite{spencer}(Fig.2 and Fig.3). This has a serious impact on the whole range of polaron parameters, in particulary to $\gamma$. Indeed, calculating of $\gamma_z$ with Eq.(7) one finds $\gamma_z=0.49$ and $\gamma_z=0.41$ for $R=$ 1 and $R=3$, respectively. These results should be compared with $\gamma_z=0.75$ and $\gamma_z=0.53$ of Ref.\cite{spencer} for the same $R=1$ and $R=3$, respectively. In this sense use of Yukawa potential provides a more mobile EHM polaron at any range of screening radius.

\section{Optical conductivity}
Optical conductivity of the lattice polarons have been studied extensively (see recent review paper \cite{asa-devr}). At strong coupling regime an optical absorption of small polarons is calculated by using generalized Einstein relation $\sigma (\nu)=eD(\nu)/\nu$, where $D(\nu)=a^2W(\nu)$ is diffusion coefficient, $W(\nu)$ is the hopping probability of the absorbtion of photon with the energy $\hbar\nu$. Explicit derivation of $\sigma(\nu)$ is performed by integrating over imaginary time and using saddle-point approximation (see for example Refs.\cite{asa-devr,bry}). The resulting formula is \cite{asa-devr}
\begin{equation}\label{14}
    \sigma(\nu)=\frac{ne^2a^2\sqrt{\pi}t^2}{2k_BT\sqrt{E_akT}}e^{-E_a/kT}\frac{\sinh\frac{\hbar\nu}{2k_BT}}{\frac{\hbar\nu}{k_BT}}e^{\hbar^2\nu^2/4\delta},
\end{equation}
where
\begin{equation}\label{15}
    E_a=\frac{k_BT}{N}\sum_{\bf q} |\gamma({\bf q})|^2[1-cos({\bf q}\cdot{\bf a})]tanh\frac{\hbar\omega_{\bf q}}{4k_BT}
\end{equation}
is the activation energy, $n$ is polarons' density, ${\bf q}$ is phonons' wave vector, $\gamma({\bf q})$ is a general form of the dimensionless electron-phonon interaction matrix element and
\begin{equation}\label{16}
    \delta=\frac{1}{2N}\sum_{\bf q} |\gamma({\bf q})|^2[1-cos({\bf q}\cdot{\bf a})]\frac{\hbar^2\omega^2_{\bf q}}{sinh(\hbar\omega_{\bf q}/2k_BT)}.
\end{equation}
The electron-phonon interaction matrix element can be written for our case as \cite{asa-b2003}
\begin{equation}\label{17}
    \gamma({\bf q})=-\frac{1}{\sqrt{M\omega^3_{\bf q}}}\sum_{\bf m}e^{-i{\bf q}\cdot{\bf m}}{\bf e}_{\bf q}\cdot\nabla_{\bf m}U_{\bf m}({\bf 0})
\end{equation}
where ${\bf e}_{\bf q}$ is phonon polarization vector. At zero temperature and when one considers single dispersionless phonon mode Eq.(14) reduces to
\begin{equation}\label{18}
\sigma(\nu)=\frac{\sigma_{0}\widetilde{t}^2}{\hbar\nu\sqrt{2E_{a}\hbar\omega}}exp\left[{-\frac{(\hbar\nu-4E_{a})^2}{(2\sqrt{2E_{a}\hbar\omega})^2}}\right],
\end{equation}
where $\sigma_{0}$ is a constant. The main difference between the optical conductivities of lattice polarons within ordinary Holstein and extended Holstein models is that in the former case an electron deforms only the site where it seats, while in the
second case it deforms also many neighboring sites. Due to the photon absorption a polaron of ordinary Holstein model hops to an undeformed site, and $E_{a}=E_{p}/2$ \cite{bry}. However a polaron of the extended Holstein model hops to a deformed neighboring site, so that $E_{a}=\gamma E_{p}/2$ \cite{asa-bratk,yavjetp}.
Extended Holstein model enables one to take into account real crystal structure of a lattice and type of an electron-lattice interaction (through parameter $\gamma$). An estimation of $\gamma$ in cuprates with long-range electron-lattice interaction force gives the value of $\simeq 0.2\div 0.3$ (see also \cite{asa-apex}).
An anomalous midinfrared optical absorption of cuprates with the band maximum energies from $0.1$ eV up to $0.5$ eV \cite{tt,or,uch,sch,bi-eklund} can be reproduced within the Extended Holstein model. Indeed, the experiment show that there are strong coupling of doped holes in the cuprates with the multiple phonon modes with the energies $27$ meV, $45$ meV, $61$ meV and $75$ meV \cite{timusk,xjzhou}. Moreover, coupling to the $75$ meV phonon mode increases with increasing of doping \cite{mcqueen}. Polaron energy estimated from the long-range density-displacement Fr\"{o}lihch-type electron-phonon interaction is found to be $\approx 0,65$ eV \cite{asa-bratk}. In accordance with these value one finds for the peak energy $E_m=2\gamma E_p$ in optical conductivity spectra $0.27$ eV and $0.39$ eV for $\gamma=0.2$ and $\gamma=0.3$, respectively.

Now let's discuss the influence of screening of an electron-phonon interaction on the optical conductivity of EHM polarons. As it was emphasized in the previous section the screened forces (Eq.(6) and Eq.(7)) affect to the all polaron parameters, in particular the polaron shift Eq.(9) and the exponent factor $\gamma$ Eq.(13). In Fig.8 an optical conductivity curves at different screening radii are presented.
\begin{figure}
\resizebox{0.5\textwidth}{!}{%
  \includegraphics{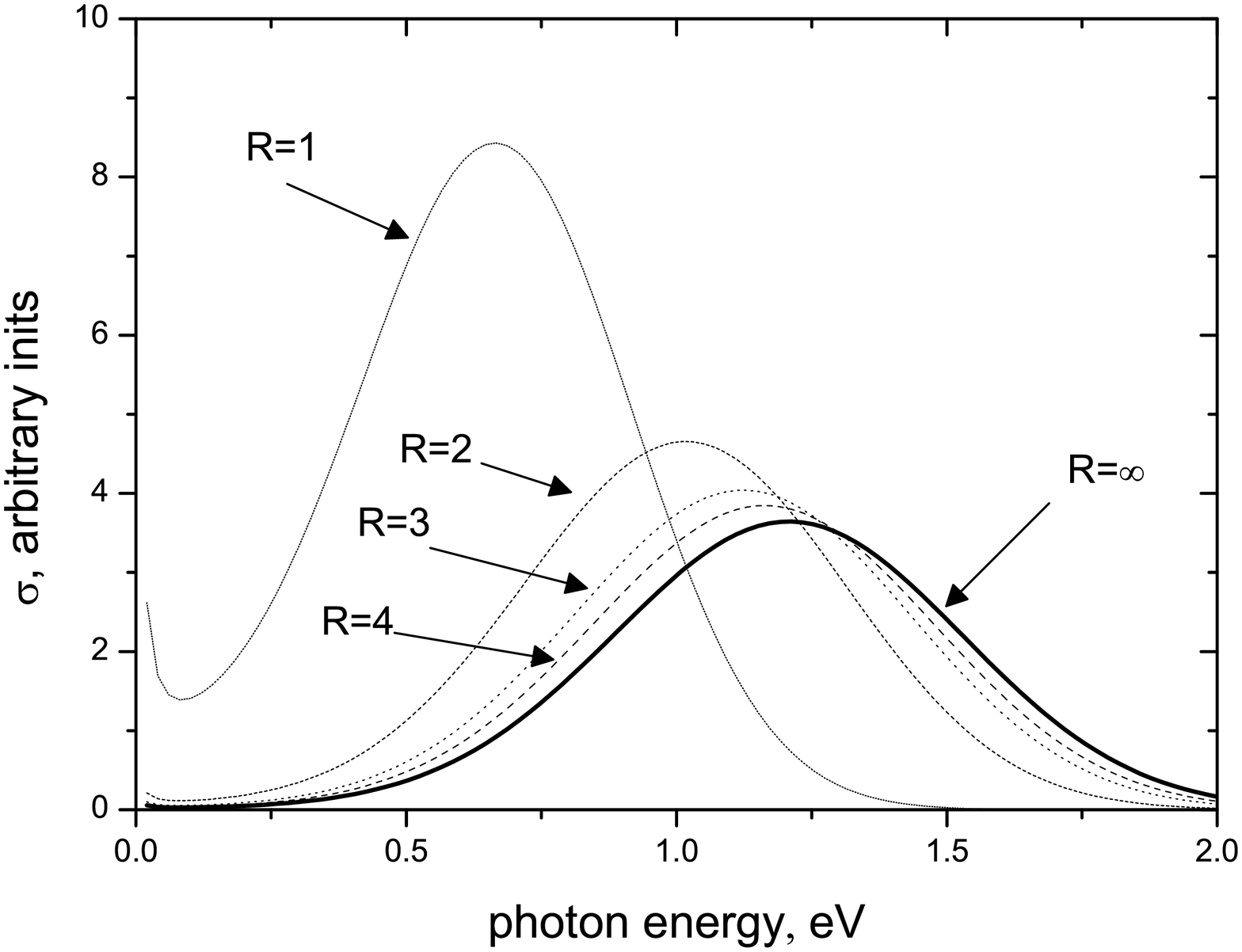}
}
\caption{Optical conductivity of EHM polarons as a function of photon energy $\hbar\nu$ at different values of screening radius $R$: $R=1$- short dotted line,
$R=2$- short dashed line, $R=3$- dotted line, $R=4$- dashed line, and $R=\infty$- solid thick line. In plotting the curves relative
change of polaron energy $E_p$ was taken into account according to Eq.(9). $\hbar\omega$=75 meV}
\label{fig:9}       
\end{figure}
\begin{figure}
\resizebox{0.5\textwidth}{!}{%
  \includegraphics{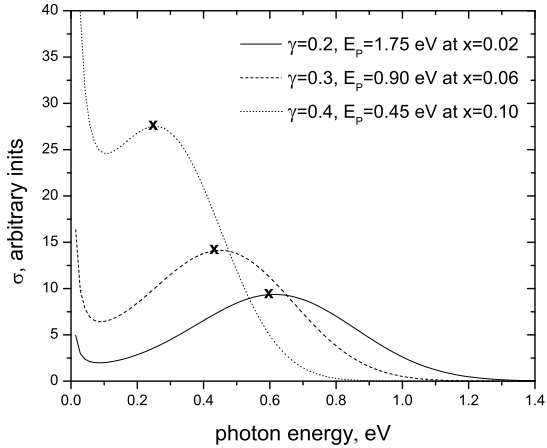}
}
\caption{Optical conductivity of EHM polarons as a function of photon energy $\hbar\nu$ for different doping levels $x$. The experimental peak energy values are shown by symbols $\times$. $\hbar\omega$=75 meV}
\label{fig:10}       
\end{figure}
At large values of screening radius $R$ compared to the lattice constant $|{\bf a}|$ the effect of screening on optical conductivity curve is not so sensitive. However, when $R\sim|{\bf a}|$ the curves are affected strongly. Screening reduce polaron shift $E_p$ and consequently an energy $E_m$. Here differently from ordinary Holstein model this reduction due to two different factors, namely $E_p$ and $\gamma$. In the ordinary Holstein model $\gamma$ always equal to $1$ and thus the reduction of the energy $E_m$ is mainly due to the reduction of $E_p$ by screening. For the extended Holstein model $\gamma\neq 1$ and it depends on screening radius $R$. As it is seen from the Fig.4 screening increases the value of $\gamma$. The polaron shift $E_p$ increases with screening radius $R$ while $\gamma$ decreases. However, the overall effect of both of them gives rise shift of $E_m$ to a lower energy values. This issue is overlooked when discussing the doping dependence of optical conductivity of cuprates in Ref. \cite{gmzhao}. Optical conductivity of our model is consistent with the results of early studies \cite{a-k-r,loos} at strong coupling limit as well as with the experiment \cite{bi-eklund}. In Ref.\cite{bi-eklund} reduction of $E_m$ in La$_{2-x}$Sr$_x$CuO$_4$ compound was reported with doping. In accordance with this report peak energy in the optical conductivity spectra $E_m$ equals to $0.6$ eV, $0.44$ eV and $0.24$ eV for doping levels ($x$)  $0.02$, $0.06$ and  $0.10$, respectively. In our model such type reduction of $E_m$ is explained by the simultaneous effect of the parameters $\gamma$ and $E_p$ on $E_m$. Optical conductivity curves of EHM polarons at different values of $\gamma$ and $E_p$  are given in Fig.10. In the same figure the experimental values of $E_m$ from Ref.\cite{bi-eklund} are shown by $\times$ symbols.
\section{Conclusion}
We have studied an effect of screening of an electron-phonon interaction on the mass and the optical conductivity of a lattice polarons within the framework of an extended Holstein model in the strong coupling regime and the nonadiabatic limit. Here screening of electron-phonon interaction is due to the presence of the excess electrons in the lattice. In order to take into account more realistic situation we have chosen a screened electron-phonon interaction potential in the form of Yukawa potential. Analytical formulas for the screened electron-phonon interaction forces are derived.  Renormalized mass of a lattice polaron and their optical conductivity are presented at different values of the screening radius. It is shown that effect of the screening on the mass and the optical conductivity of lattice polarons is more pronounced at small values of screening radius. Yukawa-type electron-phonon interaction potential provides more mobile polarons than the early studied lattice polarons \cite{kor-giant,spencer,hague-etal,hague-kor} at the same screening regime. The extended Holstein model provides to study the optical conductivity of the lattice polarons in connection with the detailed structure of a lattice and the type of electron-phonon interaction forces. Optical conductivity curves of EHM polarons are in agreement with the experimental observations.
\begin{acknowledgement}
Author is grateful to Dr M. Ermamatov for reading the \\manuscript and useful advices. This work is supported by Uzbek Academy of Science, Grant No. FA-F2-F070.
\end{acknowledgement}

%
%

\end{document}